\documentclass[12pt]{iopart}

\usepackage[compress]{cite}
\usepackage{graphicx}% Include figure files
\usepackage{bm}% bold math
\begin{document}

\title[Mean trapping time   for an arbitrary node on regular hyperbranched polymers]{Mean trapping time   for an arbitrary node on regular hyperbranched polymers}

\author{Junhao Peng $^{1,2}$ }
%\author{Junhao Peng $^{1,2}$ , Guoai Xu$^3$}
\address{1.School of Math and Information Science, Guangzhou University, Guangzhou 510006, China. \\
2.Key Laboratory of Mathematics and Interdisciplinary Sciences of Guangdong
Higher Education Institutes, Guangzhou University, Guangzhou 510006, China.
} \ead{\mailto{pengjh@gzhu.edu.cn}}

\begin{abstract}
The regular hyperbranched polymers (RHPs), also known as Vicsek fractals, are an important family of hyperbranched structures which  have attracted a wide spread attention  during the past several years.  In this paper, we study the  first-passage properties for random walks  on the RHPs. Firstly, we  propose a way to label all the different nodes of the RHPs and  derive  exact formulas to calculate the mean first-passage time (MFPT) between any two nodes and the mean trapping time (MTT) for any trap node. Then, we compare the trapping efficiency  between any two nodes of the RHPs by using the MTT as the measures of trapping efficiency. We find that the central node  of the RHPs is  the best trapping site  and the  nodes which are the farthest nodes from the  central node are  the worst trapping sites. Furthermore, we find that  the maximum of the MTT is about $4$ times more than the minimum of the MTT. The result is similar to the results in the recursive fractal scale-free trees and T-fractal, but it is quite different from that in the recursive non-fractal scale-free trees. These results can help understanding the influences of the topological properties and trap location  on  the  trapping efficiency.

\end{abstract}

%\PACS {36.20.¨Cr, 05.45.Df, 05.10.-a, 05.40.Fb,  05.60.Cd}
%05.45.Df Fractals
%\pacs{89.75.Hc}{ Networks and genealogical trees}
%\pacs{05.10.-a }{Computational methods in statistical physics and nonlinear  dynamics}
%05.40.Fb Random walks and Levy flights
%05.60.Cd Classical transport
%61.43.Hv  Fractals; macroscopic aggregates (including diffusion-limited aggregates)
%05.45.Df Fractals
%05.60.Cd Classical transport
%05.40.-a Fluctuation phenomena, random processes, noise, and Brownian motion
%89.75.Hc Networks and genealogical trees
%36.20.¨Cr Macromolecules and polymer molecules
\maketitle

\section{Introduction}
\label{intro}

In the last few decades, polymer physics has attracted considerable attention within the scientific community,
with various polymer networks proposed to describe the structures of macromolecules~\cite{GurBlu05}. Among numerous polymer networks, the regular hyperbranched polymers (RHPs), also known as Vicsek fractals,  are  important   models of the Hyperbranched polymers~\cite{GaoYan04}, which have widely applications in coatings~\cite{JoMaJaHu00, LangSten01}, conjugated functional materials~\cite{BaiZheng01, DuanQiu01}, modifiers and additives~\cite{MePlu01}, drug and gene delivery~\cite{GaoXuYan03, Uhrich97, EsTo01} etc.

In view of the widely applications of the Vicsek fractals,  interest in Vicsek fractals is growing rapidly.  Jayanthi and Wu~\cite{JayanWu92, JayanWu93, JayanWu94} succeeded in determining the eigenvalues of connectivity matrix A of the original Vicsek fractals by  determining the roots of iteratively constructed polynomials.  Blumen et al.~\cite{BlumenFer04, BlumJur03} determined the eigenvalue spectrum of general Vicsek fractals for  any generation $t$ through an algebraic iterative procedure.  From these works, one can determine the eigenvalue spectrum of very large Vicsek fractals to very high accuracy, and  then calculate many other dynamical quantities of them~\cite{BlumenVolta05a, BlumenVolta05, VoltaGa10, FuDoBl13, JurVolta11}.

Among a plethora of fundamental dynamical processes, random walks are crucial to a lot of branches of sciences and engineering and have appealed much interest~\cite{HaBe87, VoRe10, Avraham_Havlin04,  JiYang11, ma12, ChPe13, LO93, ShJaDa09}. A large variety of other dynamical processes occurring in  complex systems can be analyzed and understood in terms of random walks. Examples of these dynamics include energy or exciton transport in polymer systems~\cite{BlZu81}, reaction kinetics~\cite{BeChKl10}, and so on. A basic quantity relevant to random walks is the  mean first-passage time (MFPT) $F(x, y)$, which represents the expected number of steps for a walker starting  from the source node $x$ to arrive the trap node  $y$ for the first time. One can also define the mean trapping time (MTT) for trap node $y$ by
\begin{equation}
T_y=\frac{1}{N-1}\sum_{x\in \Omega,x\neq y}F(x,y),
\label{D_MTT}
\end{equation}
 where $\Omega$ denotes the node set and $N$ is the total number of nodes.

As is well known, the topological properties of complex system have nontrivial influences  on  the  MTT. Therefore, considerable  endeavor has been devoted to uncover the MTT for  different topological structures~\cite{ChCa99,  BeKo06, BeTuKo10,  MeAgBeVo12, ZhGuXi09, AgBu09, AgBuMa10, WuZh11, CoMi10, ZhZhGa10,  ZhWu10, ZhLiLin11, Agl08}. It is also well known that the  trap location  has great effect on the MTT and the MTT can be used as the measure of trapping efficiency for different trap node. One should analyze the MTT for an arbitrary trap node and compare the trapping efficiency among all the different traps.  The locations which  have the minimum MTT can be looked as the best trapping sites and  the locations which have  the maximum MTT  can be looked as the the  worst  trapping sites. These results have widely application in physical and chemical societies. For example, the best trap sites can be used as the best data collection sites for energy or exciton  transport in polymer~\cite{BlZu81} and geometry-controlled kinetics~\cite{BeChKl10}.

 In order to analyze the MTT  to an arbitrary trap node, one must propose a way to label all the different nodes and then derive  formulas to calculate the MTT for the different nodes. For the Cayley trees,  Zhang~\cite{LiZh13}  labeled the nodes by its levels and derived the exact analytic formula of the MTT for an arbitrary trap node. For the recursive fractal scale free trees, non-fractal scale-free trees and T-fractal, we labeled its nodes through its edge replacing  structure(i.e. the network of generation $k$, which is denoted by $G(k)$, is obtained by replacing every edge of $G(k-1)$ by a special structure)~\cite{Peng14a, Peng14b, Peng14c}. Results shows that  the ratio between the maximum and minimum of the MTT  is almost a constant  in the recursive fractal scale-free trees and T-fractal, whereas  it grows logarithmically with network order in the recursive non-fractal scale-free trees. Therefore the effect of  trap location on the MTT varies with the topological structures of the complex systems.

As for  the Vicsek fractals, they have self-similar treelike structure which  can be constructed iteratively by node replacing (i.e.  the  Vicsek fractals  of generation $k$, which is denoted by $G(k)$, is obtained by replacing every node of $G(k-1)$  with a star)~\cite{Vicsek83, BlumenFer04}. The exact analytical solution of the MTT for  the central node was obtained in Ref.~\cite{WuLiZhCh12},  the exact analytical solution of the MTT for the peripheral node  and the global mean first-passage time (i.e., the average of MFPTs over all pairs of nodes) were obtained in Ref.~\cite{LiZh13},  but the MTT for any trap node  are still unresolved and one cannot completely uncover the effect of trap location on the MTT  in the RHPs.  %Then, an interesting open question arise naturally: what is the impact of trap location on the MTT and where is the best location to trap the random walker  on the RHPs.

   Although we have proposed method to derived the exact analytic formula of the MTT for an arbitrary trap node in the recursive fractal scale-free trees and the recursive non-fractal scale-free trees~\cite{Peng14a, Peng14b}, the method  works good on  the  iterative structures obtained by  edge replacing  such as the recursive fractal  and  non-fractal scale-free trees, tree like fractal,  (u, v) flower, etc, it does not work on the  iterative structures obtained by  node replacing  such as Vicsek fractals.

   In this paper, we first propose a new way to label all the different nodes of the RHPs and  derive  exact formulas to calculate the MTT for any node. Then, we compare the trapping efficiency  between any two nodes of the RHPs and find the best and worst  trapping sites by using the MTT  as the measures of trapping efficiency. Our results show that the central node  of the RHPs is the best trapping site and  the  nodes which are the farthest nodes from the the central node are  the worst trapping sites. Finally, we find that  the maximum of the MTT is almost $\frac{3m^2+3m-2}{2m}$ times the minimum of the MTT.  The result is similar to the result in the recursive fractal scale-free trees and T-fractal, but it is quite different from that in the recursive non-fractal scale-free trees. These results can help understanding the influences of the topological properties and trap location on  the  trapping efficiency.

\section{The network model}
\label{sec:1}

\begin{figure}
\begin{center}
\includegraphics[scale=0.8]{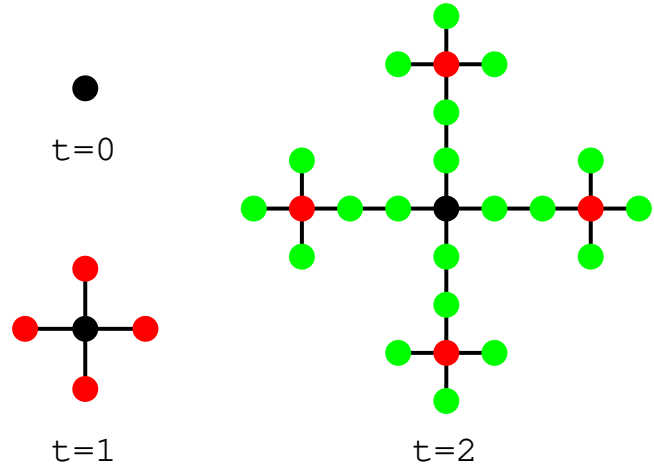}
\caption{The first three generations of the Vicsek fractals for the case $m=4$.
}
\label{structure}
\end{center}
\end{figure}

\begin{figure}
\begin{center}
\includegraphics[scale=0.8]{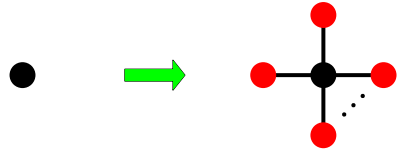}
\caption{Iterative construction method of the the Vicsek fractals, i.e., $G(t)$ is obtained from $G(t-1)$ by replacing every node of $G(t-1)$ with a star  on the right-hand side of the arrow.
}
\label{Node_rep}
\end{center}
\end{figure}

 The regular hyperbranched polymers (or Vicsek fractals)~\cite{Vicsek83, BlumJur03, BlumenFer04} of generation $t$, denoted by $G(t)$ ($t\geq 0$), are constructed in the following iterative way. For $t=0$, $G(0)$ consists of an isolated node without any edge. For $t=1$, $m$ $(m\geq 2)$ new nodes are generated with each being connected to the node of $G(0)$ to form $G(1)$, which is exactly a star. For $t\geq2$, $G(t)$ is obtained from $G(t-1)$. The detailed process is as follow. We introduce $m$ new identical copies of $G(t-1)$ and arrange them around the periphery of the original $G(t-1)$, and add $m$ new edges, each of them connecting a peripheral node in one of the $m$ corner copy structures and a peripheral node of the original central structure, where a peripheral node is a node farthest from the central node.  The first three generations of the Vicsek fractals for the case $m=4$ are shown in figure~\ref{structure}. The  Vicsek fractals $G(t)$ can also be constructed by another method, i.e., $G(t)$ is obtained from $G(t-1)$ by replacing every node of $G(t-1)$ with a star as shown in figure~\ref{Node_rep}.

According to its construction,  at each generation the total number of the nodes increases by a factor  $m+1$; therefore, the total number of nodes of $G(t)$ is $N_t=(m+1)^t$, and the total number of edges of $G(t)$ is $E_t=N_t-1=(m+1)^t-1$.

%%%%%%%%%%%%%%%%%%%%%%%%%%%%%%%%%%%%%%%%%%%%%%%%%%%%%%%%%
% Figure  1
%%%%%%%%%%%%%%%%%%%%%%%%%%%%%%%%%%%%%%%%%%%%%%%%%%%%%%%%%%
%%%%%%%%%%%%%%%%%%%%%%%%%%%%%%%%%%%%%%%%%%%%%%%%%%%%%%%%%%

%%%%%%%%%%%%%%%%%%%%%%%%%%%%%%%%%%%%%%%%%%%%%%%%%%%%%%%%%
% Figure  2
%%%%%%%%%%%%%%%%%%%%%%%%%%%%%%%%%%%%%%%%%%%%%%%%%%%%%%%%%%
%%%%%%%%%%%%%%%%%%%%%%%%%%%%%%%%%%%%%%%%%%%%%%%%%%%%%%%%%%

\section{The MTT  for random walks on Vicsek fractals}

\label{sec:det_meth}
\subsection{Simplification of the expressions for the MTT}
\label{sec:gen_meth}

For any two nodes $x$ and $y$ of  Vicsek fractals $G(t)$, $F(x,y)$ is  the MFPT from  $x$ to $y$, the sum
$$k(x,y)=F(x,y)+F(y,x)$$
is called the commute time and the MFPT can be expressed in term  of commute times~\cite{Te91}:
\begin{equation}
F(x,y)\!=\!\frac{1}{2}\!\left\{\!k(x,y)\!+\!\!\sum_{u\in G(t)}\!\pi(u)[k(y,u)-k(x,u)]\!\right\}\!,
\label{FXY}
\end{equation}
where \lq\lq$u\in G(t)$\rq\rq means that $u$ belongs to the nodes set of $G(t)$, $\pi(u)=\frac{d_u}{2E_t}$ is the stationary distribution for random walks on  Vicsek fractals  and $d_u$ is the degree of node $u$.

If we view the networks under consideration as electrical networks  by considering each edge to be a unit resistor and let $\Psi_{xy}$ denote the effective resistance  between two nodes $x$ and $y$ in the electrical networks,
 we have~\cite{Te91}
\begin{equation}
k(x,y)=2E_t\Psi_{xy},
\label{KR}
\end{equation}
 where $E_t$ is the total numbers of edges of $G(t)$. Since the  Vicsek fractals we study  are trees, the effective resistance between any two nodes is just the  shortest-path length  between the two nodes. Hence
 \begin{equation}
 \Psi_{xy}=L_{xy},
 \end{equation}
 where $L_{xy}$ denote  the shortest path length between node $x$ to node $y$.  Thus
 \begin{equation}
k(x,y)=2E_tL_{xy}.
\label{KL}
\end{equation}
Replacing $k(x,y)$ from Eq.~(\ref{KL}) in Eq.~(\ref{FXY}), and defining
\begin{equation}
S_x= \sum_{y \in G(t)}{L_{xy}},%\frac{1}{N_t-1}\cdot
\label{SX}
\end{equation}
\begin{equation}
W_x=\sum_{u\in G(t)}\pi(u)L_{xu}=\frac{1}{2E_t}\cdot \sum_{u \in G(t)}{(L_{xu}\cdot d_u)},
\label{WY}
\end{equation}
\begin{equation}
\Sigma=\sum_{u\in G(t)}\left(\pi(u)\sum_{x\in G(t)}L_{xu}\right),
\label{WS}
\end{equation}
we obtain
\begin{eqnarray}
F(x,y)%&=&E_t\!\left\{\!L_{xy}\!+\!\!\sum_{u\in G(t)}\!\pi(u)L_{yu}\!-\!\!\sum_{u\in G(t)}\!\pi(u)L_{xu} \!\right\}\! \nonumber\\
      &=&E_t(L_{xy}+W_y-W_x).
\label{FXYL}
\end{eqnarray}
Substituting  $F(x,y)$   from Eq.~(\ref{FXYL}) in Eqs.~(\ref{D_MTT}), one  gets
\begin{eqnarray} \label{MTT}
T_y%&=&\frac{1}{E_t}\sum_{x\in G(t),x\neq y}F(x,y)\nonumber  \\
%&=&\sum_{x\in G(t),x\neq y}\left(L_{xy}\!+\!\sum_{u\in G(t)}\pi(u)L_{yu}\!-\!\sum_{u\in G(t)}\pi(u)L_{xu} \right)  \nonumber  \\
%&=&\sum_{x\in G(t),x\neq y}L_{xy}+\sum_{x\in G(t),x\neq y}\sum_{u\in G(t)}\pi(u)L_{yu}-\sum_{x\in G(t),x\neq y}\sum_{u\in G(t)}\pi(u)L_{xu} \nonumber  \\
%&=&\sum_{x\in G(t)}L_{xy}+N_t\sum_{u\in G(t)}\pi(u)L_{yu}-\sum_{x\in G(t)}\sum_{u\in G(t)}\pi(u)L_{xu}   \nonumber  \\
%&=&\sum_{x\in G(t)}L_{xy}+N_t\sum_{u\in G(t)}\pi(u)L_{yu}-\Sigma
&=&S_y+N_t\cdot W_y-\Sigma.
\end{eqnarray}

Hence, if we can calculate $\Sigma$ and $S_y, W_y$ for any node $y$, we can calculate $F(x,y)$ for any two nodes $(x,y)$ and  the MTT for any  node $y$. In this paper, we  calculate these quantities of the RHPs  based on its self-similar structure.

 \subsection{General methods of calculating  the MTT}
 \label{Met_SW}
 According to the construction of  Vicsek fractals,  $G(t)$ is composed of $m+1$ copies,  called subunit,  of $G(t-1)$ which are connected  with each other by their peripheral nodes. For convenience,  we classify the subunits of $G(t)$ into different levels and let $\Lambda_k$ denote the subunit of level $k$ $(k\geq0)$. In this paper, $G(t)$ is said to be subunit of level $0$. For any   $k\geq0$, the $m+1$  subunits of $\Lambda_k$ are said to be  subunits of level $k+1$. Thus, any node of $G(t)$  is a subunit of level $t$ and $\Lambda_k$ is a copy of  Vicsek fractals with generation $t-k$. Similarity, we classify all the nodes of $G(t)$ into different levels  and the node which is the central node of certain subunit $\Lambda_k$ $(k\geq0)$ is said to belong to level $k$. The reason for we only assign  level $k$ to the central node of subunit $\Lambda_k$ is we can use the same labels to label the subunit $\Lambda_k$ and its  central node.

\begin{figure}
\begin{center}
\includegraphics[scale=0.7]{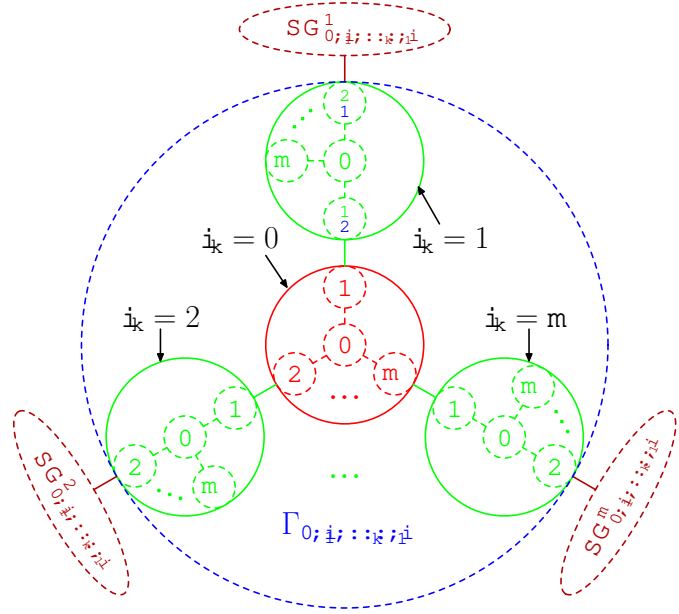}
\caption{Construction of  subunit   $\Gamma_{0, i_1, i_2, ..., i_{k-1}}$ and the way we label its subunits. It is represented by the biggest dashed circle and is composed of $m+1$ subunits $\Gamma_{0, i_1, i_2, ..., i_{k}}$ ($i_k=0,1,2, ..., m$) represented by solid circles. It  connects with other part of $G(t)$ (i.e., $SG^{i_k}_{0,i_1,\cdots,i_{k-1}}, i_k=1,2, ..., m$) at its $m$ corners. Each subunit  $\Gamma_{0, i_1, i_2, ..., i_{k}}$ is also  composed of $m+1$ subunits $\Gamma_{0, i_1, i_2, ..., i_{k+1}}$ represented by small dashed circles.  The value of $i_k$  (or the numbers in every small dashed circles) shows the relation between  $i_k$ (or $i_{k+1}$) and the locations of the corresponding subunits in  subunit  $\Gamma_{0, i_1, i_2, ..., i_{k-1}}$.}
\label{sub_k}       % Give a unique label
\end{center}
\end{figure}

 In order to distinguish the subunits of different locations, inspired by  the method of Ref~\cite{MeAgBeVo12}, we label the subunit  $\Lambda_k$ $(0\leq k \leq t)$ by a sequence $\{0, i_1, i_2, ..., i_{k} \}$ and denote it by $\Gamma_{0, i_1, i_2, ..., i_{k}}$, where $i_j=0, 1, 2, \cdots, m$ $(1\leq j \leq k)$ labels its location in its parent subunit $\Gamma_{0, i_1, i_2, ..., i_{k-1}}$. In particular,  \textquoteleft $\{0\}$\textquoteright  represents the Vicsek fractals $G(t)$ itself. figure~\ref{sub_k} shows the construction of $\Gamma_{0, i_1, i_2, ..., i_{k-1}}$ and the way we label its subunits.
 As shown in figure~\ref{sub_k},  $\Gamma_{0, i_1, i_2, ..., i_{k-1}}$ $(k>0)$, which is represented by the biggest dashed circle,  is composed of $m+1$ subunits $\Gamma_{0, i_1, i_2, ..., i_{k}}$ ($i_k=0,1,2, ..., m$) represented by solid circles. It also  connects with other part of $G(t)$ (i.e., $SG^{i_k}_{0,i_1,\cdots,i_{k-1}}, i_k=1,2, ..., m$) at its $m$ corners.  Each subunit  $\Gamma_{0, i_1, i_2, ..., i_{k}}$ is also  composed of $m+1$ subunits $\Gamma_{0, i_1, i_2, ..., i_{k+1}}$ represented by small dashed circles. We label the subunit  at the center of $\Gamma_{0, i_1, i_2, ..., i_{k-1}}$ by $i_k=0$ and the $m$ peripheral subunits surround the cental one by $i_k=1,2,\cdots, m$. The value of $i_k$  shows the relation between  $i_k$ and the location of  subunit $\Gamma_{0, i_1, i_2, ..., i_{k}}$  in  subunit $\Gamma_{0, i_1, i_2, ..., i_{k-1}}$.  The numbers in each small dashed circles, which are the corresponding values of $i_{k+1}$,  show the relation between $i_{k+1}$ and the location of  subunit $\Gamma_{0, i_1, i_2, ..., i_{k+1}}$  in $\Gamma_{0, i_1, i_2, ..., i_{k}}$ and $\Gamma_{0, i_1, i_2, ..., i_{k-1}}$. But there are two numbers in the two small dashed circles for $i_k=1$. it means the way to label the two subunits of $\Gamma_{0, i_1, i_2, ..., i_{k-1}}$ should be divided into two cases. If $\Gamma_{0, i_1, i_2, ..., i_{k-1}}$ is the central subunit of $G(t)$ (i.e., $i_j=0$ for $j=1,2,\cdots,k-1$), the dashed circle  near the center of $\Gamma_{0, i_1, i_2, ..., i_{k-1},1}$ should be labeled by $i_{k+1}=1$, the other one should be labeled by $i_{k+1}=2$; otherwise, the labels for the two subunits should be exchanged.  According to the way we label the subunits, for any subunit $\Gamma_{0, i_1, i_2, ..., i_{k-1}}$ $(k\geq 1)$,  we find
 \begin{equation}
    N^{1}_{0,i_1, ..., i_{k\!-\!1}} \!\geq \! N^{2}_{0,i_1,  ..., i_{k\!-\!1}}\!\geq \! \cdots \!\geq \! N^{m}_{0,i_1, ..., i_{k\!-\!1}},
 \label{Com_N_ik}
 \end{equation}
 %$N^{1}_{0,i_1, i_2, ..., i_{k-1}}\geq N^{2}_{0,i_1, i_2, ..., i_{k-1}}\geq \cdots \geq N^{m}_{0,i_1, i_2, ..., i_{k-1}}$,
  where $N^{i_k}_{0,i_1, i_2, ..., i_{k-1}}$ denote the total numbers of nodes of $SG^{i_k}_{0,i_1,\cdots,i_{k-1}}$  $(i_k=1,2,\cdots,m)$. The calculation of $N^{i_k}_{0,i_1, i_2, ..., i_{k-1}}$ and the proof of Eq.~(\ref{Com_N_ik}) are presented in  ~\ref{N_ik} and  ~\ref{PRoof_N_ik} respectively. %It implies that $i_{k}=1$ represented the subunit $\Gamma_{0, i_1, i_2, ..., i_{k}}$ which is the nearest subunit to the central node of $G(t)$ among all subunits of level $k$ in $\Gamma_{0, i_1, i_2, ..., i_{k-1}}$.

%  In the $m$ peripheral subunits  $\Gamma_{0, i_1, i_2, ..., i_{k}}$ $(i_k=1,2,\cdots,m)$, $i_{k}=1$ represented the subunit $\Gamma_{0, i_1, i_2, ..., i_{k}}$ which is the nearest subunit to the central node of $G(t)$ among all subunits of level $k$ in $\Gamma_{0, i_1, i_2, ..., i_{k-1}}$, $i_{k}=2$ represented the subunit $\Gamma_{0, i_1, i_2, ..., i_{k}}$  which is connected with  $SG^{i_k}_{0,i_1,\cdots,i_{k-1}}$.

 %%%%%%%%%%%%%%%%%%%%%%%%%%%%%%%%%%%%%%%%%%%%%%%%%%%%%%%%%
% Figure  3
%%%%%%%%%%%%%%%%%%%%%%%%%%%%%%%%%%%%%%%%%%%%%%%%%%%%%%%%%%
%%%%%%%%%%%%%%%%%%%%%%%%%%%%%%%%%%%%%%%%%%%%%%%%%%%%%%%%%%

For any  node $x\in G(t)$, it must be a central node of certain subunit $\Gamma_{0, i_1, i_2, ..., i_{k}}$ (note: for any terminal node of $G(t)$, it can be viewed as a subunit of level $t$ which has only one node, then it can also be regarded as the central node of this subunit). For convenience, we also  label the node $x$ by the  same sequence $\{0, i_1, i_2, ..., i_{k} \}$. Therefore we can use this label to represent  \textquotedblleft $x$\textquotedblright   in symbol \textquotedblleft $S_x$\textquotedblright,  \textquotedblleft $W_x$\textquotedblright, \textquotedblleft $T_x$\textquotedblright and \textquotedblleft $D_x$\textquotedblright.
%, where  \textquoteleft $\{0\}$\textquoteright  represents the center node of $G(t)$
  As derived in  ~\ref{SxWx}, for any $k\geq1$,
\begin{eqnarray}
S_{\{0,i_1, i_2, ..., i_{k} \}}&=&S_{\{0,i_1, i_2, ..., i_{k-1} \}}+3^{t-k}\left[(m+1)^t \right.\nonumber \\
   & &\left.-2(m+1)^{t-k}-2N^{i_k}_{0,i_1, i_2, ..., i_{k-1}} \right],\quad
\label{RSX}
\end{eqnarray}
and
\begin{eqnarray}
W_{\{0,i_1, i_2, ..., i_{k} \}}&\!=\!&W_{\{0,i_1, i_2, ..., i_{k-1} \}}\!+\!\frac{3^{t\!-\!k}(m\!+\!1)^t}{E_t}\!\left[\!(m\!+\!1)^t \right.\nonumber \\
   & &\!\left.\!\!-\!2(m\!+\!1)^{t-k}\!-\!2N^{i_k}_{0,i_1, i_2, ..., i_{k-1}}\! \right]\!,
\label{RWX}
\end{eqnarray}
where
\begin{equation}
N^{0}_{0,i_1, i_2, ..., i_{k-1}}=[(m+1)^t-2(m+1)^{t-k}]/2,
\label{N0k}
\end{equation}
%$N^{0}_{0,i_1, i_2, ..., i_{k-1}}=[(m+1)^t-2(m+1)^{t-k}]/2$
 and $N^{i_k}_{0,i_1, i_2, ..., i_{k-1}}$ $(i_k=1,2,\cdots,m)$, which denote the total numbers of nodes of  $SG^{i_k}_{0,i_1,\cdots,i_{k-1}}$ $(i_k=1,2,\cdots,m)$,  are calculated in  ~\ref{N_ik}.

 %As for $S_{\{0\}}$ and $W_{\{0\}}$, we have  derived them in  Appendix~\ref{S0_W0}.
 Replacing $S_x$ and $W_x$  with  the right-hand side of Eqs.~(\ref{RSX}) and (\ref{RWX})  in Eqs.~(\ref{MTT}), we obtain the MTT  for  node  labeled as  $\{0, i_1, i_2, ..., i_{k} \}$:
 %For any $k\geq1$, as derived in Appendix~\ref{SxWx},
\begin{eqnarray}
T_{\{0,i_1, i_2, ..., i_{k} \}}&=&T_{\{0,i_1, i_2, ..., i_{k-1} \}}\!+\!3^{t\!-\!k}\frac{2E_t+1}{E_t}\!\left[\!(m\!+\!1)^t \right.\nonumber \\
   & &\!\left.\!\!-\!2(m\!+\!1)^{t-k}\!-\!2N^{i_k}_{0,i_1, i_2, ..., i_{k-1}}\! \right]\!.
\label{RTX}
\end{eqnarray}
Using Eq.~(\ref{RTX}) repeatedly, we obtain
\begin{eqnarray}
&&T_{\{0,i_1, i_2, ..., i_{k} \}}\nonumber \\
&=&T_{\{0\}}\!+\!\frac{2E_t+1}{E_t}\!\left\{\!(m\!+\!1)^t\sum_{j=1}^k{3^{t-j}}\!-\!2\sum_{j=1}^k(3m\!+\!3)^{t-j} \right.\nonumber \\
   & &\!\left.\!\!-\!2\sum_{j=1}^k3^{t-j}N^{i_j}_{0,i_1, i_2, ..., i_{j-1}}\! \right\}\!
   \nonumber \\
   &=&T_{\{0\}}\!+\!\frac{2E_t\!+\!1}{2E_t}\!\left\{\!(m\!+\!1)^t(3^t-3^{t\!-\!k})\!-\!\frac{4}{3m+2}\times \right.\nonumber \\
   & &[{(3m\!+\!3)^{t}\!-\!(3m\!+\!3)^{t\!-\!k}}]\!-\!4\sum_{j=1}^k3^{t\!-\!j}N^{i_j}_{0,i_1, i_2, ..., i_{j\!-\!1}}\! \}\!.
\label{TX}
\end{eqnarray}

As for $T_{\{0\}}$, we have calculated them as  examples in Sec.~\ref{sec:example}. Therefore, we can calculate the MTT  for any node.

 \subsection{Examples}
\label{sec:example}
In order to explain our methods,   we calculate the  MTT  for node   labeled as $\{0\}$ and nodes denoted by $P_k$ $(1\leq k\leq t)$ with labels $\{0,i_1, i_2, ..., i_{k}\}=\{0, \underbrace{m, m,\cdots, m}_k\}$. They are the farthest nodes from the central node $\{0\}$ among all nodes of level $k$.

For node  labeled as $\{0\}$, inserting Eqs.~(\ref{S0} ), (\ref{W0} ) and (\ref {Xigmat} )  into Eqs.~(\ref{MTT}), we obtain
\begin{eqnarray}
%\begin{equation}
\label{MTT0}
T_{\{0\}}
%=\frac{[4m(m+1)^{t-1}-4][(3m+3)^{t}-1]-(3m^2-4m-4)(m+1)^{t-1}(3^t-1)}{[(m+1)^t-1]*(6m+4)},
&=&\frac{1}{[(m+1)^t-1]*(6m+4)}\left\{ 4m 3^t(m+1)^{2t-1}\right.\nonumber  \\
&&\left.\!-\!(m\!+\!1)^{t\!-\!1} (m^23^{t\!+\!1}\!-\!3m^2\!+\!8m\!+\!4)\!+\!4\right\}.
%\end{equation}
\end{eqnarray}
The result   is consistent with that  derived in Ref.~\cite{WuLiZhCh12}.

For any nodes  $P_k$ $(1\leq k\leq t)$, note that $N^{m}_{0,\underbrace{m, m,\cdots, m}_k}=0$ and $E_t=(m+1)^t-1$. Let $\{0,i_1, i_2, ..., i_{k}\}=\{0, \underbrace{m, m,\cdots, m}_k\}$, replace $T_{\{0\}}$  from Eqs.~(\ref{MTT0})   in Eqs.~(\ref{TX}), one gets
\begin{eqnarray}
%\begin{equation}
\label{MTTPk}
T_{P_k}
&=&\frac{1}{[(m\!+\!1)^t\!-\!1](6m\!+\!4)}\left\{ (m\!+\!1)^{2t\!-\!1}[(6m^2\!+\!6m\!-\!4)\right.\nonumber  \\
&&\times 3^t\!-\!(6m^2\!+\!10m\!+\!4)3^{t\!-\!k}]\!+\!8\times 3^{t\!-\!k}(m\!+\!1)^{2t\!-\!k} \nonumber  \\
&&\!-\!(m\!+\!1)^{t\!-\!1}[(6m^2\!+\!m\!-\!2) 3^t\!-\!(3m^2\!+\!5m\!+\!2)3^{t\!-\!k}\nonumber  \\
&&\left.\!-\!3m^2\!+\!8m\!+\!4]\!-\!4\times (3m\!+\!3)^{t\!-\!k}\!+\!4\right\}.
%\end{equation}
\end{eqnarray}

%%%%%%%%%%%%%%%%%%%%%%%%%%%%%%%%%%%%%%%%%%%%%%%%%%%%%%%%%%
%% Figure  4
%%%%%%%%%%%%%%%%%%%%%%%%%%%%%%%%%%%%%%%%%%%%%%%%%%%%%%%%%%%
\begin{figure}
 \begin{center}
    \includegraphics[scale=0.4]{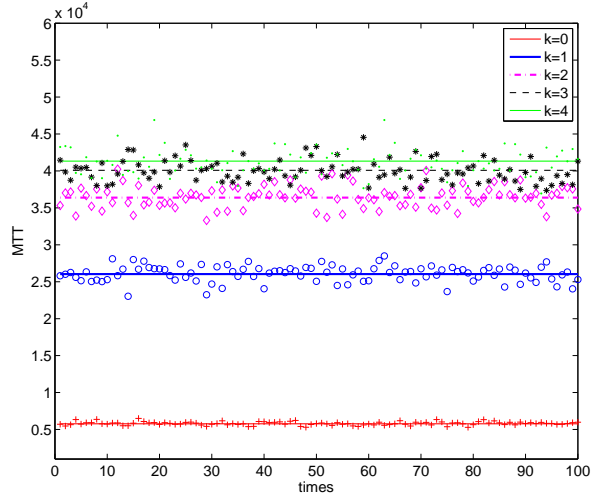}
  \caption{The MTT for nodes $P_k$ $(k=0, 1, 2, 3, 4)$.}
  \label{Fig_MTTs}
  \end{center}
\end{figure}
%%%%%%%%%%%%%%%%%%%%%%%%%%%%%%%%%%%%%%%%%%%%%%%%%%%%%%%%%%
These results  are consistent with those obtained by simulation  we have just done. The comparison between simulation results and derived results for  nodes $P_k$ $(k=0, 1, 2, 3, 4)$ in Vicsek fractals with $m=4$, $t=4$ are shown in figure~\ref{Fig_MTTs}.  The horizontal axis stands for the different times, the vertical axis is the MTT, the lines with different shape and color stand for the  derived results, the scattered dots with the same color represent the corresponding results obtained at  different time's simulation. Averaging the $100$ times'  results and comparing them with the derived results, we find the relative error is less than $10^{-3}$.

\section{Effect of trap location on trapping efficiency in Vicsek fractals}% Maximum and Minimum of MTT and MDT
\label{sec:Com_MTT_MDT}

In this section,  we  compare the trapping efficiency  among all the nodes of Vicsek fractals by using the MTT  as the measures of trapping efficiency, and then find the best and the worst  trapping sites. %the best trapping sites(i.e., nodes which have the minimum MTT) and the  worst trapping sites(i.e., nodes which have the maximum MTT).
%we first compare the MTT for nodes in the same subunit $\Gamma_{0, i_1, i_2, ..., i_{k-1}}$ $(k=1,2,\cdots,t)$, then
 Because any node of $G(t)$  is in one to one correspondence with a sequence  $\{0, i_1,\cdots,i_t\}$, we can find from Eqs.~(\ref{TX}) that the difference of the MTT  for different nodes  depends on $\sum_{j=1}^t3^{t\!-\!j}N^{i_j}_{0,i_1, i_2, ..., i_{j\!-\!1}}$, and that nodes with maximum MTT must have the minimum $\sum_{j=1}^t3^{t\!-\!j}N^{i_j}_{0,i_1, i_2, ..., i_{j\!-\!1}}$ , whereas  nodes with minimum MTT must have the maximum $\sum_{j=1}^t3^{t\!-\!j}N^{i_j}_{0,i_1, i_2, ..., i_{j\!-\!1}}$.

In order to find the maximum and minimum $\sum_{j=1}^t3^{t\!-\!j}N^{i_j}_{0,i_1, i_2, ..., i_{j\!-\!1}}$ among all nodes of $G(t)$, we  compare $N^{i_k}_{0,i_1, i_2, ..., i_{k-1}}$ ($i_k=0,1,2,\cdots,t$) in any fixed subunit $\Gamma_{0, i_1, i_2, ..., i_{k-1}}$ $(k=1,2,\cdots,t)$, the results can be divided into two case.

 Case I: If $\Gamma_{0, i_1, i_2, ..., i_{k-1}}$ is the central subunit of $G(t)$ (i.e., $\{i_1, i_2, ..., i_{k-1}\}=\{0, 0, \cdots, 0\}$), we obtain  %$N^{i_k}_{0,i_1, i_2, ..., i_{k-1}}=\frac{(m+1)^t-(m+1)^{t-k}}{m}$;
   \begin{equation}
 \left\{
  \begin{array}{l}
   N^{0}_{0,i_1, ..., i_{k-1}}\!>\!N^{1}_{0,i_1, ..., i_{k\!-\!1}}   \\
   N^{1}_{0,i_1, ..., i_{k\!-\!1}} \!= \! N^{2}_{0,i_1,  ..., i_{k\!-\!1}}\!= \! \cdots \!= \! N^{m}_{0,i_1, ..., i_{k\!-\!1}}
  \end{array}
  \right.
 \label{ComNk1}
 \end{equation}
 by comparing  Eq.~(\ref{N0k}) with Eq.~(\ref{N_00k}).
 %\begin{equation*}
%  \begin{array}{l}
%  N^{i_k}_{0,i_1, i_2, ..., i_{k\!-\!1}}\!<\!N^{0}_{0,i_1, i_2, ..., i_{k\!-\!1}},\quad i_k=1,2,\cdots, m. \\
%  N^{i_k}_{0,i_1, i_2, ..., i_{k\!-\!1}}=N^{j_k}_{0,i_1, i_2, ..., i_{k\!-\!1}},\quad i_k,j_k=1,2,\cdots, m.
%  \end{array}
% \label{ComNk1}
% \end{equation*}
% $N^{1}_{0,i_1, i_2, ..., i_{k-1}}=N^{2}_{0,i_1, i_2, ..., i_{k-1}}=\cdots=N^{m}_{0,i_1, i_2, ..., i_{k-1}}$$<N^{0}_{0,i_1, i_2, ..., i_{k-1}}$.

 Case II: If $\{i_1, i_2, ..., i_{k-1}\}\neq \{\underbrace{0, 0, \cdots, 0}_{k-1}\}$, as derived in  ~\ref{PRoof_N_ik},
 %ccording to the way we label the subunit,
  \begin{equation}
    N^{1}_{0,i_1, ..., i_{k\!-\!1}} \!\geq \! N^{2}_{0,i_1,  ..., i_{k\!-\!1}}\!\geq \! \cdots \!\geq \! N^{m}_{0,i_1, ..., i_{k\!-\!1}}.
 \label{ComNk2}
 \end{equation}
 Therefore, the central node of $G(t)$ must be a node of $SG^{1}_{0,i_1,\cdots,i_{k-1}}$  which have the maximum  number of nodes among all the subgraphs $SG^{i_k}_{0,i_1,\cdots,i_{k-1}}$ $(i_k=1,2,\cdots,m)$. Hence,
  \begin{equation}
 N^{1}_{0,i_1, ..., i_{k\!-\!1}} \!>\! \frac{(m+1)^t}{2}\!>\!N^{0}_{0,i_1, ..., i_{k\!-\!1}}.
 \end{equation}
 Note that $\sum_{k=1}^mN^{0}_{0,i_1, ..., i_{k\!-\!1}}=(m+1)^t-(m+1)^{t-k-1}$. we obtain
   \begin{equation}
   \label{Nmax}
 N^{1}_{0,i_1, ..., i_{k\!-\!1}} \!>\!N^{0}_{0,i_1, ..., i_{k\!-\!1}}\!>\!N^{2}_{0,i_1, ..., i_{k\!-\!1}}.
 \end{equation}
%   \begin{equation}
% \left\{
%  \begin{array}{l}
%   N^{1}_{0,i_1, ..., i_{k\!-\!1}}\!>\!N^{0}_{0,i_1, ..., i_{k-1}}\!>\!N^{2}_{0,i_1, ..., i_{k\!-\!1}}   \\
%   N^{2}_{0,i_1, ..., i_{k\!-\!1}} \!\geq \! N^{3}_{0,i_1,  ..., i_{k\!-\!1}}\!\geq \! \cdots \!\geq \! N^{m}_{0,i_1, ..., i_{k\!-\!1}}
%  \end{array}
%  \right..
% \label{ComNk2}
% \end{equation}
 Therefore, for  nodes with label $\{0, i_1,\cdots,i_t\}$, let $i_1=i_2=\cdots=i_t=m$ (i.e., the node $P_t$ of Sec.~\ref{sec:example}). We find from Eqs.~(\ref{ComNk1}),(\ref{ComNk2}) and (\ref{Nmax}) that it  has the minimum $\sum_{j=1}^t3^{t\!-\!j}N^{i_j}_{0,i_1, i_2, ..., i_{j\!-\!1}}$ among all nodes of $G(t)$. Thus, %it has the maximum MTT and it is the  worst trapping site.
   \begin{equation}
   %T_{max}=
   T_{\{0,\underbrace{m,m,\cdots,m}_t\}}\!= \! max \{T_x: x\in G(t)\}.
 \label{Tmax}
 \end{equation}

Note that any node of $G(t)$ (except the central node with label $\{0\}$) can be labeled by $\{\underbrace{0, 0, \cdots 0}_k, i_k, i_{k+1},\cdots,i_t\}$ $(i_k\neq 0, k=1,2,\cdots, t)$, let $i_{k+1}=i_{k+2}=\cdots=i_t=1$. We find from Eqs.~(\ref{ComNk1}),(\ref{ComNk2}) and (\ref{Nmax}) that it  has the maximum $\sum_{j=1}^t3^{t\!-\!j}N^{i_j}_{0,i_1, i_2, ..., i_{j\!-\!1}}$ among all these kind of nodes. That is to say, for any $i_k\neq 0$ $( k=1,2,\cdots, t)$ and  $i_j=0,1,2,\cdots,m$ $(t\geq j>k)$,
%Let $\Omega_k\equiv \{T_{\underbrace{0, 0, \cdots 0}_k, i_k, i_{k+1},\cdots,i_t}: i_j=0,1,2,\cdots,t, , t\geq j>k\}$, we have,
  \begin{equation}
   T_{\{\underbrace{0, 0, \cdots 0}_k, i_k, \underbrace{1,\cdots,1}_{t-k}\}}\!\leq \! T_{\{\underbrace{0, 0, \cdots 0}_k, i_k, i_{k+1},\cdots,i_t\}}.
 \label{TminCk}
 \end{equation}
%holds for all $i_j=0,1,2,\cdots,m$ $(t\geq j>k)$ and $i_k\neq 0$.

As proved in in  ~\ref{sec:Proof_of_TminCk}, for any $k=1,2,\cdots, t$ and $i_k\neq 0$,
  \begin{equation}
   T_{\{0\}}\!< \! T_{\{\underbrace{0, 0, \cdots 0}_k, i_k, \underbrace{1,\cdots,1}_{t-k-1}\}}.
 \label{Tmin0}
 \end{equation}
Eqs.~(\ref{TminCk}) and (\ref{Tmin0}) imply
   \begin{equation}
   %T_{min}=
  T_{\{0\}}\!= \! min \{T_x: x\in G(t)\}.
 \label{Tmin}
 \end{equation}

Let $k=t$ in Eq.~(\ref{MTTPk}) and compare it with $T_0$ shown in Eq.~(\ref{MTT0}), while $t\rightarrow \infty$,
   \begin{equation}
   %\frac{T_{max}}{T_{min}}=
   \frac{T_{\{0,\underbrace{m,m,\cdots,m}_t\}}}{T_{\{0\}}}\!\approx\! \frac{3m^2+3m-2}{2m}>4.
 \label{Rmax_min}
 \end{equation}
%which implies that there is a big difference between the maximum  and minimum of the MTT,  thus the trap's position has big effect on the trapping efficiency.
Comparing the result with that in the recursive fractal or non-fractal scale-free trees~\cite{Peng14a, Peng14b}, we find that  the effect of  trap location on the MTT in the RHPs  is similar to the result in the recursive fractal scale-free trees, but it is quite different from that in the recursive non-fractal scale-free trees.

\section{Conclusion}
\label{sec:4}

Firstly,  a way to label  the  nodes of the RHPs is proposed in this paper. It is inspired by the  method of Ref~\cite{MeAgBeVo12}. Although  the  method of Ref~\cite{MeAgBeVo12}  has broadly application and   works good on  the  iterative structures obtained by  edge replacing
 %(i.e. the network of generation $k$, which is denoted by $G(k)$, is obtained by replacing every edge of $G(k-1)$ by a special structure)
 , such as tree like fractal,  (u, v) flower, etc, it does not work on the  iterative structures obtained by  node replacing %(i.e. the network  $G(k)$ is obtained by replacing every node of $G(k-1)$ by a special structure)
 , such as Vicsek fractals. Our method  works good on Vicsek fractals and it is also suitable for other   iterative structures obtained by  node replacing.

Then, we derive  formulas  to calculate the MTT  for any  node and compare the trapping efficiency for any two nodes of the RHPs by using the MTT as the measures of trapping efficiency. Our results show that the central node  of the RHPs is the best trapping site and  the  nodes which are the farthest nodes from the the central node are  the worst trapping sites.  One can find the direct applications of the results, e.g., if we  study  energy or  exciton transport on the RHPs, our results show that the central node  is the best data collection site.

Finally, we find that the ratio between the maximum and minimum of the MTT in RHPs is almost a constant. The result is similar to the result in the recursive fractal scale-free trees and T-fractal, but it is quite different from that in the recursive non-fractal scale-free trees which grows logarithmically with network order. What are the reasons for the difference and  what are the results for other networks? They are still   interesting unresolved problems.

Having  the MFPT and the MTT for unbiased random walks on unweighted RHPs, some further works might be the MFPT and the MTT for biased random walks on weighted (or unweighted) RHPs~\cite{LinZhang14a, PengZhang14, LinZhang14b}. Although the method we calculate the MFPT and MTT does not work directly on this case, the method we label the nodes of RHPs is still suitable for this case and the relation between the commute time and effective resistance is also an useful bridge.

%But the MFPT is just the first moment for the distribution of first passage time (FPT),  it  is not a sufficient measure to characterize the first passage dynamics of a system. Having obtained the MFPT for any pair of nodes, one can  further analyze the distribution of FPT. Explicitly deriving the distribution of the FPT is still  an interesting unresolved problem.

\ack{
The authors are grateful to the anonymous referees for their valuable comments and suggestions. This work was supported  by
the scientific research program of Guangzhou municipal colleges and universities under Grant No. 2012A022 and the research program of Guangzhou Education Science "Twelfth Five-Year Plan" under Grant No. 12A030.
}

%\appendix %{Appendix}
\appendix
\section{Calculation of $N^{i_k}_{0,i_1, i_2, ..., i_{k-1}}$ }
\label{N_ik}
For any subunit  $\Gamma_{0, i_1, i_2, ..., i_{k-1}}$ $(k\geq 1)$, $N^{i_k}_{0,i_1, i_2, ..., i_{k-1}}$  $(i_k=1,2,\cdots,m)$ denote the total numbers of nodes of subgraph $SG_{i_k}$ which is connected with $\Gamma_{0, i_1, i_2, ..., i_{k}}$, as shown in figure~\ref{sub_k}. For $k=1$, note that $\Gamma_{0}$ is $G(t)$ itself and there is no node surround $\Gamma_{0}$, therefore,
  $$N^{i_1}_{0}=0, \qquad i_1=1,2,\cdots,m.$$

 Assuming that $N^{i_k}_{0,i_1, i_2, ..., i_{k-1}}$  $(i_k=1,2,\cdots,m$, $k\geq 1)$ are known, we now analyze $N^{i_{k+1}}_{0,i_1, i_2, ..., i_{k}}$. Note that the total number of nodes for subunit $\Gamma_{0, i_1, i_2, ..., i_{k}}$ is $(m+1)^{t-k}$, if $i_{k}=0$ (see the central red solid circle in figure~\ref{sub_k}), for any $i_{k+1}=1,2,\cdots,m$,
 \begin{equation}
N^{i_{k+1}}_{0,i_1, i_2, ..., i_{k-1},0}=N^{i_{k+1}}_{0,i_1, i_2, ..., i_{k-1}}+(m+1)^{t-k}.
\label{N_0}
\end{equation}

 If $i_{k}\neq0$ (see the  green solid circles in figure~\ref{sub_k}), the calculation is divided into two cases.

 Case I:  $\Gamma_{0, i_1, i_2, ..., i_{k-1}}$ is the central subunit of $G(t)$ (i.e., $i_j=0$ for $j=1,2,\cdots,k-1$),
for any  $i_k=1,2,\cdots, m$ and $i_{k+1}=1,2,\cdots, m$, %F we find
  \begin{eqnarray} \label{N_K1}
&&\!N^{i_{k+1}}_{0,i_1, i_2, ..., i_{k-1},i_k}\! \nonumber\\
%&=&
%\!\left \{ \!                %×óÀ¨ºÅ
%  \begin{array}{ll}
%   \!\sum_{j\neq i_{k}}N^{j}_{0,i_1, i_2, ..., i_{k-1}}\! & i_{k+1}=1 \\
%   N^{i_{k}}_{0,i_1, i_2, ..., i_{k-1}} & i_{k+1}=2 \\
%   0 & i_{k+1}>2
%  \end{array}
%\right.
%\nonumber\\
 &\!=\!&
\!\left \{ \!                %×óÀ¨ºÅ
  \begin{array}{ll}
   (m\!+\!1)^t\!-\!(m\!+\!1)^{t\!-\!k}\!-\!N^{i_{k}}_{\!0,i_1, i_2, ..., i_{k-1}\!} & i_{k\!+\!1}\!=\!1 \\
   N^{i_{k+1}}_{0,i_1, i_2, ..., i_{k-1}} & i_{k\!+\!1}\!=\!2 \\
   0 &  i_{k\!+\!1}\!>\!2
  \end{array}
\right..\quad
\end{eqnarray}

Case II: If $\Gamma_{0, i_1, i_2, ..., i_{k-1}}$ is not the central subunit of $G(t)$,  for any  $i_{k+1}=1,2,\cdots,m$,
 \begin{eqnarray} \label{N_K2}
&&\!N^{i_{k+1}}_{0,i_1, i_2, ..., i_{k-1}, 1}\! \nonumber\\
%&=&
%\!\left \{ \!                %×óÀ¨ºÅ
%  \begin{array}{ll}
%   N^{1}_{0,i_1, i_2, ..., i_{k-1}} & i_{k+1}=1 \\
%   \!\sum_{j\neq i_{k}}N^{j}_{0,i_1, i_2, ..., i_{k-1}}\! & i_{k+1}=2 \\
%   0 & i_{k+1}>2
%  \end{array}
%\right.
%\nonumber\\
&\!=\!&
\!\left \{ \!                %×óÀ¨ºÅ
  \begin{array}{ll}
  N^{i_{k+1}}_{0,i_1, i_2, ..., i_{k-1}} & i_{k\!+\!1}\!=\!1 \\
  (m\!+\!1)^t\!-\!(m\!+\!1)^{t\!-\!k}\!-\!N^{i_{k}}_{\!0,i_1, i_2, ..., i_{k-1}\!}& i_{k\!+\!1}\!=\!2 \\
   0 &  i_{k\!+\!1}\!>\!2
  \end{array}
\right.,\quad
\end{eqnarray}
 and for $i_k=2,3,\cdots, m$ , $i_{k+1}=1,2,\cdots,m$,
\begin{eqnarray} \label{N_K3}
&&\!N^{i_{k+1}}_{0,i_1, i_2, ..., i_{k-1},i_k}\! \nonumber\\
%&=&
%\!\left \{ \!                %×óÀ¨ºÅ
%  \begin{array}{ll}
%   \!\sum_{j\neq i_{k}}N^{j}_{0,i_1, i_2, ..., i_{k-1}}\! & i_{k+1}=1 \\
%   N^{i_{k}}_{0,i_1, i_2, ..., i_{k-1}} & i_{k+1}=2 \\
%   0 & i_{k+1}>2
%  \end{array}
%\right.
%\nonumber\\
&\!=\!&
\!\left \{ \!                %×óÀ¨ºÅ
  \begin{array}{ll}
   (m\!+\!1)^t\!-\!(m\!+\!1)^{t\!-\!k}\!-\!N^{i_{k}}_{\!0,i_1, i_2, ..., i_{k-1}\!} & i_{k\!+\!1}\!=\!1 \\
   N^{i_{k+1}}_{0,i_1, i_2, ..., i_{k-1}} & i_{k\!+\!1}\!=\!2 \\
   0 &  i_{k\!+\!1}\!>\!2
  \end{array}
\right..\quad
\end{eqnarray}

Therefore, we can calculate  $N^{i_k}_{0,i_1, i_2, ..., i_{k-1}}$ $(i_k=1,2,\cdots,m)$ for any subunit  $\Gamma_{0, i_1, i_2, ..., i_{k-1}}$ $(k\geq 1)$.
%it is straightforward that let $\xi_1$
%They have been derived in Appendix~\ref{N_ik}.

For example, if $\{0, i_1, i_2, ..., i_{t}\}=\{0, \underbrace{m, m, ..., m}_t\}$,
 \begin{equation}
 N^{m}_{0,\underbrace{m, m, ...,m}_k}=0, \quad k=0, 1, \cdots, t-1.
\label{N_mmk}
\end{equation}

If $\{0, i_1, i_2, ..., i_{t}\}=\{\underbrace{0, 0, ..., 0}_k, i_k, \underbrace{1, 1, ..., 1}_{t-k} \}$  $(k\geq 1$, $i_k\neq 0)$, using Eq.~(\ref{N_0}) repeatedly, we obtain
 \begin{equation}
 N^{i_{k}}_{\underbrace{0, 0, ...,0}_k}=\frac{(m\!+\!1)^t-(m\!+\!1)^{t\!-\!k\!+\!1}}{m}
\label{N_00k}
\end{equation}
Let $i_{k+1}=1$ in Eq.~(\ref{N_K1}), we get
 \begin{eqnarray}
 N^{1}_{\underbrace{0, 0, ...,0}_k, i_{k}}&=&(m\!+\!1)^t\!-\!(m\!+\!1)^{t\!-\!k}\!-\!N^{i_{k}}_{\underbrace{0, 0, ...,0}_k}\nonumber \\
 &=&\frac{(m\!-\!1)(m\!+\!1)^t+(m\!+\!1)^{t\!-\!k}}{m}
\label{N_k11}
\end{eqnarray}
We can also obtain from Eq.~(\ref{N_K2}) that, for any  $j$ $(j=1, 2, \cdots, t-k-1)$,
\begin{eqnarray}
N^{1}_{\underbrace{0, 0, ...,0}_k, i_{k},\underbrace{1, 1, ...,1}_j }= N^{1}_{\underbrace{0, 0, ...,0}_k, i_{k}}.
\label{N_k11j}
\end{eqnarray}

\section{Proof of Eq.~(\ref{Com_N_ik})}
\label{PRoof_N_ik}
We prove Eq.~(\ref{Com_N_ik}) by mathematical induction.

Step 1: For $k=1$,  Eq.~(\ref{Com_N_ik}) is true for subunit $\Gamma_{0}$, since $N^{i_{1}}_{0}=0$ for any $i_1=1,2, \cdots, m$.

Step 2: Suppose Eq.~(\ref{Com_N_ik}) is true for any  subunit $\Gamma_{0, i_1, i_2, ..., i_{k-1}}$ with some $(k\geq1)$. % with  $k=l\geq 1$.
%that is
% \begin{equation}
%    N^{1}_{0,i_1, ..., i_{l\!-\!1}} \!\geq \! N^{2}_{0,i_1,  ..., i_{l\!-\!1}}\!\geq \! \cdots \!\geq \! N^{m}_{0,i_1, ..., i_{l\!-\!1}}.
% \label{Com_N_il}
% \end{equation}%with  $k=l+1$.
Then we prove it also hold for $k+1$, that is to say,
  \begin{equation}
    N^{1}_{0,i_1, ..., i_{k\!-\!1},i_k} \!\geq \! N^{2}_{0,i_1,  ..., i_{k\!-\!1},i_k }\!\geq \! \cdots \!\geq \! N^{m}_{0,i_1, ..., i_{k\!-\!1},i_k}
 \label{Com_N0-m}
 \end{equation}
 is also true for any  subunit $\Gamma_{0, i_1, i_2, ..., i_{k}}$ $(i_k=0,1,2,\cdots,m)$). %with  $k=l+1$ (i.e., $\Gamma_{0, i_1, i_2, ..., i_{l}}$

If $i_{k}=0$, we obtain Eq.~(\ref{Com_N0-m}) from Eq.~(\ref{N_0}) and the induction hypothesis.
% \begin{equation}
%    N^{1}_{0,i_1, ..., i_{k\!-\!1},0} \!\geq \! N^{2}_{0,i_1,  ..., i_{k\!-\!1},0}\!\geq \! \cdots \!\geq \! N^{m}_{0,i_1, ..., i_{k\!-\!1},0}.
% \label{Com_N0}
% \end{equation}
%for  subunit $\Gamma_{0, i_1, i_2, ..., i_{k-1},0}$

If $i_{k}\neq0$, The proof is divided into two cases.

 Case I:  $\Gamma_{0, i_1, i_2, ..., i_{k-1}}$ is the central subunit of $G(t)$ (i.e., $i_j=0$ for $j=1,2,\cdots,k-1$), according to Eq.~(\ref{N_K1}),   we find, for any $i_k=1,2,\cdots,m$,
 \begin{equation}
 N^{i_{k+1}}_{0,i_1, i_2, ..., i_{k}}=0, \quad   i_{k+1}=3,\cdots,m,
  \end{equation}
 and $$N^{2}_{0,i_1, i_2, ..., i_{k-1},i_k}=N^{i_{k+1}}_{0,i_1, i_2, ..., i_{k-1}}<\frac{N_t}{2}.$$

 Note that $SG^{1}_{0, i_1, i_2, ..., i_{k}}$ is the subgraph containing the central node of $G(t)$ (see figure~\ref{sub_k}). Therefore,
  $$N^{1}_{0,i_1, i_2, ..., i_{k-1},i_k}>\frac{N_t}{2}.$$
 Hence, Eq.~(\ref{Com_N0-m}) holds for any $i_k=1,2,\cdots,m$.
%  \begin{equation}
%    N^{1}_{0,i_1, ..., i_{k\!-\!1},i_k} \!\geq \! N^{2}_{0,i_1,  ..., i_{k\!-\!1},i_k }\!\geq \! \cdots \!\geq \! N^{m}_{0,i_1, ..., i_{k\!-\!1},i_k}.
% \label{Com_N1-m}
% \end{equation}

Case II:  $\Gamma_{0, i_1, i_2, ..., i_{k-1}}$ is not the central subunit of $G(t)$. $SG^{1}_{0, i_1, i_2, ..., i_{k-1}}$ must be the subgraph containing the central node of $G(t)$ (see figure~\ref{sub_k}). Therefore,  for any $i_k=1,2,\cdots,m$,
  $$N^{1}_{0,i_1, i_2, ..., i_{k-1}}>\frac{N_t}{2}.$$
Hence,
 $$N^{1}_{0,i_1, i_2, ..., i_{k}}\geq N^{1}_{0,i_1, i_2, ..., i_{k-1}}>\frac{N_t}{2}.$$
 But  for any $i_k=1,2,\cdots,m$,
 \begin{equation}
 N^{i_{k+1}}_{0,i_1, i_2, ..., i_{k}}=0, \quad   i_{k+1}=3,\cdots,m.
  \end{equation}
  Thus, Eq.~(\ref{Com_N0-m}) holds for any $i_k=1,2,\cdots,m$.

 \section{Derivation of Eqs.~(\ref{RSX}) and (\ref{RWX})}
\label{SxWx}
%\section{Derivation of $S_{\{0, i_1, i_2, ..., i_{k} \}}$ and $W_{\{0, i_1, i_2, ..., i_{k} \}}$ }
%\label{SxWx}
%Therefore $\Gamma_{0, i_1, i_2, ..., i_{k-1}}$ is the parent subunit of subunit $\Gamma_{i_1, i_2, ..., i_{k}}$.
For any node $x$ of Vicsek fractals   labeled  by $\{0, i_1, i_2, ..., i_{k} \}$, $0\leq i_j \leq m$, $j=1,2,..., k$, $k\geq1$, it is the central node of subunit  $\Gamma_{0, i_1, i_2, ..., i_{k}}$.  If $i_k=0$, $\Gamma_{0, i_1, i_2, ..., i_{k}}$ and $\Gamma_{0, i_1, i_2, ..., i_{k-1}}$ have the same central node. Thus,
\begin{eqnarray}
S_{\{0, i_1, i_2, ..., i_{k-1}, 0 \}}&=&S_{\{0, i_1, i_2, ..., i_{n-1} \}}, \\
W_{\{0, i_1, i_2, ..., i_{k-1}, 0 \}}&=&W_{\{0, i_1, i_2, ..., i_{n-1} \}}.
\label{RSX0}
\end{eqnarray}
If we denote  $N^{0}_{0,i_1, i_2, ..., i_{k-1}}=[(m+1)^t-2(m+1)^{t-k}]/2$, it is straightforward that Eqs.~(\ref{RSX}) and (\ref{RWX}) hold for $i_k=0$.

If $i_k=1$, as shown in figure~\ref{sub_k}, $\Gamma_{0, i_1, i_2, ..., i_{k-1},1}$ connects with $SG^1_{0,i_1,\cdots,i_{k-1}}$ by an edge and connects with other part of Vicsek fractals by $\Gamma_{0, i_1, i_2, ..., i_{k-1},0}$. Both $\Gamma_{0, i_1, i_2, ..., i_{k-1},0}$ and $\Gamma_{i_1, i_2, ..., i_{k-1},1}$ are copies of Vicsek fractals of generation $t-k$. We denote by $p$, the node which is labeled by $\{0, i_1, i_2, ..., i_{k-1} \}$. Node $p$ is also the central nodes of subunit  $\Gamma_{0, i_1, i_2, ..., i_{k-1},0}$. By symmetry, we have
\begin{eqnarray}
\sum_{y\in G_{0}}L_{xy}=\sum_{y\in G_{1}}L_{py},\\
\sum_{y\in G_{1}}L_{xy}=\sum_{y\in G_{0}}L_{py},
\end{eqnarray}
and
\begin{eqnarray}
\sum_{y\in G_{0},G_{1}}\pi(y)L_{xy}=\sum_{y\in G_{0}, G_{1}}\pi(y)L_{py}+(m-1)\frac{3^{t-k}}{2E_t}. \nonumber
\end{eqnarray}
where $G_{0}$ , $G_{1}$  are the simplifications of $\Gamma_{0,i_1, i_2, ..., i_{k-1},0}$ and $\Gamma_{0,i_1, i_2, ..., i_{k-1},1}$ respectively.
 Let $G_{others}$ denote the rest part of Vicsek fractals except for $SG^1_{0,i_1,\cdots,i_{k-1}}$, $\Gamma_{0,i_1, i_2, ..., i_{k-1},0}$ and $\Gamma_{0,i_1, i_2, ..., i_{k-1},1}$, the total numbers of nodes of $G_{others}$  is $(m+1)^t-2(m+1)^{t-k}-N^{1}_{0,i_1, i_2, ..., i_{k-1}}$. We find that for any node $y\in G_{others}$,  $L_{xy}=L_{py}+L_{px}$ and that for any node $y\in SG^1_{0,i_1,\cdots,i_{k-1}}$,  $L_{xu}=L_{py}-L_{px}$. Hence,%S_{\{0, i_1, i_2, ..., i_{n-1} \}}
\begin{eqnarray}
&&S_{\{0, i_1, i_2, ..., i_{k-1},1 \}}=S_x= \sum_{y \in G(t)}{L_{xy}}  \nonumber \\
&=&\sum_{y \in SG_{1}}{L_{xy}}+\sum_{y \in G_{0}}{L_{xy}}+\sum_{y \in G_{1}}{L_{xy}}+ \sum_{y \in G_{others}}{L_{xy}}\nonumber \\
&=&\sum_{y \in SG_{1}}{(L_{py}-L_{xp})}+\sum_{y \in G_{0}}{L_{py}}+\sum_{y \in G_{1}}{L_{py}}\nonumber \\
&&+ \sum_{y \in G_{others}}{(L_{py}+L_{xp})}\nonumber \\
&=&\sum_{y \in G(t)}{L_{py}}+L_{xp}\left[(m+1)^t-2(m+1)^{t-k}-2N^{1}_{i_{k-1}} \right]  \nonumber \\
&=&S_p+3^{t-k}\left[(m+1)^t-2(m+1)^{t-k}-2N^{1}_{ i_{k-1}} \right].
\end{eqnarray}
where  $N^{1}_{i_{k-1}}$ and $SG_{1}$ are the simplifications of $N^{1}_{0,i_1, i_2, ..., i_{k-1}}$ and $SG^1_{0,i_1,\cdots,i_{k-1}}$ respectively.  Therefore, Eq.~(\ref{RSX})  holds for $i_k=1$.

Similarity,
\begin{eqnarray}
&&W_{\{0, i_1, i_2, ..., i_{k-1},1 \}}=W_x= \sum_{y \in G(t)}{\pi(y)L_{xy}}  \nonumber \\
&=&\sum_{y \in SG_{1}}{\pi(y)L_{xy}}+\sum_{y \in G_{0},G_{1}}{\pi(y)L_{xy}}+ \sum_{y \in G_{others}}{\pi(y)L_{xy}}\nonumber \\
&=&\sum_{y \in SG_{1}}{\pi(y)(L_{py}-L_{xp})}+ \sum_{y \in G_{others}}{\pi(y)(L_{py}+L_{xp})}\nonumber \\
&&+\sum_{y \in G_{0},G_{1}}{\pi(y)L_{py}}+(m-1)\frac{3^{t-k}}{2E_t}\nonumber \\
%&=&\sum_{y \in G(t)}{\pi(y)L_{py}}+\frac{3^{t-k}}{E_t}\left[(m+1)^t-2(m+1)^{t-k}-2N^{1}_{i_{k-1}} \right]  \nonumber \\
&=&W_p+\frac{3^{t-k}}{E_t}\left[(m+1)^t-2(m+1)^{t-k}-2N^{1}_{i_{k-1}} \right].
\end{eqnarray}
Therefore, Eq.~(\ref{RWX})  holds for $i_k=1$.

By symmetry, we can also verify that Eqs.~(\ref{RSX}) and (\ref{RWX}) hold for $i_k=2,3,\cdots,m$.

 \section{Derivation of $S_{\{0 \}}$ and $W_{\{0 \}}$}%Eqs.~(\ref{S0}) and (\ref{W0})
\label{S0_W0}
%Let $P_k$ $(0\leq k\leq t)$ denote  the node whose labels $\{0,i_1, i_2, ..., i_{k}\}=\{0,2, 2, ..., 2\}$.
In this section, we derive $S_x$ and $W_x$ for node $x$ labeled by $\{0 \}$, which is the central node of $G(t)$. But it is difficult to calculate them directly. We first calculate $S_x$ and $W_x$ for node denoted by $P$, which is the farthest node to the central node $\{0 \}$. Then we calculate $S_{\{0 \}}$ and $W_{\{0 \}}$ from Eqs.~(\ref{RSX}) and (\ref{RWX}). %with label $\{0,i_1, i_2, ..., i_{t}\}=\{0, \underbrace{m, m,\cdots, m}_t\}$.

In order to tell the difference of $S_{P}$ and $W_{P}$  for Vicsek fractals of different generation $t$ $(0\leq t)$,
 we denote by $S_{P}^t$, $W_{P}^t$ the $S_{P}$ and $W_{P}$ in Vicsek fractals of generation $t$ respectively. it is straightforward that $S_{P}^0=0$ and $W_{P}^0=0$. For $t>0$, according  to the self-similar structure, $S_{P}^t$  satisfies the following recursion relation:
\begin{eqnarray}
S_{P}^t&=&S_{P}^{t-1}+[S_{P}^{t-1}+N_{t-1}3^{t-1}]  \nonumber \\%(1+L_{t-1})
&&+(m-1)\cdot[S_{P}^{t-1}+2N_{t-1}3^{t-1}].  \nonumber \\%(2+2L_{t-1})
&=&(m+1) S_{P}^{t-1}+(2m-1)(m+1)^{t-1}3^{t-1}.
\label{RPt}
\end{eqnarray}
Using Eq.~(\ref{RPt}) repeatedly, we obtain
\begin{eqnarray}
S_{P} &\equiv &S_{P}^t =(m+1) S_{P}^{t-1}+(2m-1)(m+1)^{t-1}3^{t-1}  \nonumber \\
      &=&(m+1)^2 S_{P}^{t-2}+(2m-1)(m+1)^{t-1}(3^{t-2}+3^{t-1}) \nonumber \\
      &=&\cdots \nonumber \\
      &=&(m\!+\!1)^t S_{P}^{0}+(2m\!-\!1)(m\!+\!1)^{t\!-\!1}(1\!+\!3^1\!+\!\cdots+3^{t\!-\!1})  \nonumber \\
      &=&(2m\!-\!1)(m\!+\!1)^{t\!-\!1}\frac{3^t-1}{2}.
      \label{SP}
\end{eqnarray}
Similarity,
%%derivation of recurision relation of WPt
%\begin{eqnarray}
%W_{P}^t&=&\frac{2E_{t-1}}{2E_t}\left\{(W_{P}^{t-1}+\frac{L_{t-1}}{2E_{t-1}})+[W_{P}^{t-1}\right. \nonumber \\%(1+L_{t-1})
%&&+\frac{(m-1)L_{t-1}}{2E_{t-1}} +(1+L_{t-1})(1+\frac{m}{2E_{t-1}})] \nonumber \\
%&&\left. +(m-1)[W_{P}^{t-1}+(2+2L_{t-1})(1+\frac{1}{2E_{t-1}})]\right\}.
%\label{RWt}
%\end{eqnarray}
\begin{eqnarray}
W_{P} &\equiv &W_{P}^t =\frac{(m\!+\!1)^{t\!-\!1}\!-\!1}{(m\!+\!1)^{t}\!-\!1}(m\!+\!1) W_{P}^{t\!-\!1}\!-\!\frac{m}{2[(m\!+\!1)^{t}\!-\!1]}\nonumber \\
      &&+3^{t-1}(2m-1)\frac{(m\!+\!1)^{t\!-\!1}}{(m\!+\!1)^{t}\!-\!1} \nonumber \\
      &=&\cdots \nonumber \\
      &=&0\cdot W_{P}^{0}+\frac{(2m\!-\!1)(m\!+\!1)^{t\!-\!1}}{(m\!+\!1)^{t}\!-\!1}(1\!+\!3^1\!+\!\cdots+3^{t\!-\!1}) \nonumber \\
      && -\frac{m}{2[(m\!+\!1)^{t}\!-\!1]} [1\!+\!(m\!+\!1)^1\!+\!\cdots+(m\!+\!1)^{t\!-\!1}]\nonumber \\
      &=& \frac{1}{2[(m\!+\!1)^{t}\!-\!1]}\{(m\!+\!1)^{t\!-\!1}\![(\!2m\!-\!1\!)\!3^t\!-\!3m\!]+\!1\!\}\!.
      \label{WP}
\end{eqnarray}
Let $P_k$ $(0\leq k\leq t)$ denote  the node whose label satisfies $\{0,i_1, i_2, \cdots, i_{k}\}=\{0, \underbrace{m, m,\cdots, m}_k\}$. We have  $P_t\equiv P$ and $P_0\equiv \{0\}$. Note that $N^{m}_{0,\underbrace{m, m,\cdots, m}_k}=0$, for any $k$ $(1\leq k\leq t)$. Therefore, we can obtain from Eqs.~(\ref{RSX}) and (\ref{RWX}) that
\begin{eqnarray}
S_{P_{k-1}}&=&S_{P_{k}}-3^{t-k}\left[(m+1)^t-2(m+1)^{t-k} \right],\nonumber
\label{RSPK}
\end{eqnarray}
\begin{eqnarray}
W_{P_{k\!-\!1}}&\!=\!&W_{P_{k}}\!-\!\frac{3^{t-k}}{2[(m\!+\!1)^{t}\!-\!1]}\left[2(m+1)^t\!-\!4(m+1)^{t\!-\!k} \right]. \nonumber
\label{RWPK}
\end{eqnarray}
Thus
\begin{eqnarray}
S_{\{0\}}&\equiv &S_{P_0}=S_{P_{1}}-3^{t-1}\left[(m+1)^t-2(m+1)^{t-1} \right]\nonumber \\
&=&S_{P_t}-(m+1)^t[1+3^1+\cdots+3^{t-1}]\nonumber \\
&&+2[1+(3m+3)^1+\cdots+(3m+3)^{t-1} ]\nonumber \\
&\!=\!&(m\!-\!2)(m\!+\!1)^{t\!-\!1}\frac{3^t\!-\!1}{2}\!+\!2\cdot\frac{(3m\!+\!3)^{t}\!-\!1}{3m\!+\!2},
\label{S0}
\end{eqnarray}
and
\begin{eqnarray}
&&W_{\{0\}}\equiv W_{P_0}%\!=\!W_{P_{1}}\!-\!\frac{3^{t-1}}{2[(m\!+\!1)^{t}\!-\!1]}\left[2(m+1)^t\!-\!4(m+1)^{t\!-\!1} \right]
\nonumber \\
&=&\!W_{P_t}\!-\!\frac{2(m+1)^t}{2[(m\!+\!1)^{t}\!-\!1]}(1+3^1+\cdots+3^{t-1})\nonumber \\
&&\!+\!\frac{4}{2[(m\!+\!1)^{t}\!-\!1]}[1\!+\!(3m\!+\!3)^1\!+\!\cdots\!+\!(3m\!+\!3)^{t\!-\!1} ]\nonumber \\
&\!=\!& \frac{1}{2[(m\!+\!1)^{t}\!-\!1]}\left\{ (m\!+\!1)^{t\!-\!1}[(m\!-\!2)3^t-2m+1]\right.\nonumber \\
&&\left.+1+4\cdot\frac{(3m\!+\!3)^{t}\!-\!1}{3m\!+\!2}\right\}.
\label{W0}
\end{eqnarray}

 \section{Exact calculation of $\Sigma$ }
\label{SSWP}
We find that
 $$\Sigma=\sum_{u\in G(t)}(\pi(u)\sum_{x\in G(t)}L_{xu})=\sum_{u\in G(t)}W_u.$$
 Because any node of $G(t)$  is in one to one correspondence with a sequence  $\{0, i_1,\cdots,i_t\}$, Thus
\begin{equation}
 \Sigma=\sum_{i_1,\cdots,i_t}W_{\{0, i_1,\cdots,i_t\}}.
  \label{xigma}
\end{equation}
where the  summation run over all the possible values of $i_k=0,1,2,\cdots, m$ ($1\leq k \leq t$).
For any $k$ ($0\leq k\leq t$), let
\begin{equation}
 \Sigma_k=\sum_{i_1,\cdots,i_k}W_{\{0, i_1,\cdots,i_k\}}.
  \label{xigma}
\end{equation}
Therefore $\Sigma_0=W_{\{0\}}$. Note that
\begin{eqnarray}
 \sum_{i_k=0}^m N^{i_k}_{\!0,i_1, i_2, ..., i_{k-1}\!}\!=\! \frac{3}{2}(m\!+\!1)^t\!-\!(m\!+\!1)^{t-k+1} \!+\!(m\!+\!1)^{t-k}.\nonumber
\label{SumNk}
\end{eqnarray}
For any $k$ ($1\leq k\leq t$), replacing $W_{\{0, i_1,\cdots,i_k\}}$ from Eq.~(\ref{RWX}) in Eq.~(\ref{xigma}), we obtain%\right.
\begin{eqnarray}
 \Sigma_k&=&\sum_{i_1,\cdots,i_{k-1}}\sum_{i_k=0}^m W_{\{0, i_1,\cdots,i_k\}}
\nonumber \\
&=&\sum_{i_1,\cdots,i_{k-1}}\sum_{i_k=0}^m \left\{ W_{\{0,i_1, i_2, ..., i_{k-1} \}}\!+\!\frac{3^{t\!-\!k}}{E_t}\right.\nonumber \\
   & &\left.\times \!\left[\!(m\!+\!1)^t\!-\!2(m\!+\!1)^{t-k}\!-\!2N^{i_k}_{0,i_1, i_2, ..., i_{k-1}}\! \right]\right\}\!\nonumber \\
&=&(m+1)\Sigma_{k-1}\!+\!(m+1)^{k-1}\frac{3^{t\!-\!k}}{E_t}\nonumber \\
   & &\times \!\left[\!(m-2)(m\!+\!1)^t\!+\!2(m\!+\!1)^{t-k} \right]\!.
\label{RXigmaK}
\end{eqnarray}
Using equation (\ref{RXigmaK}) repeatedly and replacing $\Sigma_{0}$ with $W_{\{0\}}$ (see Eq.~(\ref{W0})), we obtain
\begin{eqnarray}
 \Sigma&\equiv &\Sigma_t=(m+1)\Sigma_{t-1}\!+\!(m+1)^{t-1}\frac{3^{0}}{E_t}\nonumber \\
   & &\times \!\left[\!(m-2)(m\!+\!1)^t\!+\!2(m\!+\!1)^{0} \right]\! \nonumber \\
   &=&(m\!+\!1)^t\Sigma_{0}\!+\!\frac{(m\!+\!1)^{2t-1}}{E_t}(m\!-\!2)[1+3^{1}\!+\!\cdots\!+\!3^{t\!-\!1}]\nonumber \\
   & &\!+\!\frac{2(m\!+\!1)^{t\!-\!1}}{E_t}[1\!+\!(3m\!+\!3)^{1}\!+\!\cdots\!+\!(3m\!+\!3)^{t\!-\!1}]\!\nonumber \\
   &=&\frac{1}{2E_t}\left\{(m\!+\!1)^{2t\!-\!1}[2(m\!-\!2)3^t\!-\!3m\!-\!3]\!+\!(m\!+\!1)^{t}\right.\nonumber \\
   & &\left.\!+\!\frac{(3m\!+\!3)^{t}-1}{3m\!+\!2}(4m\!+\!8)(m\!+\!1)^{t\!-\!1}\right\},
\label{Xigmat}
\end{eqnarray}
where $E_t=(m+1)^t-1$.
\section{Proof of Eq.~(\ref{TminCk})  }
\label{sec:Proof_of_TminCk}
For any node with label $\{0,i_1, i_2, ..., i_{t}\}$, Let $k=t$ in Eq.~(\ref{TX}), we can obtain
\begin{eqnarray}
&&T_{\{0,i_1, i_2, ..., i_{t}\}}-T_{\{0\}}=\frac{2E_t\!+\!1}{2E_t}\Phi(i_1, i_2, ..., i_{t}),
\label{Sub_TX}
\end{eqnarray}
where
\begin{eqnarray}
\Phi(i_1, i_2, ..., i_{t})&=&(m\!+\!1)^t(3^t-1)\!-\!\frac{4}{3m+2}[(3m\!+\!3)^{t} \!-\!1] \nonumber \\
  &&\!-\!4\sum_{j=1}^t3^{t\!-\!j}N^{i_j}_{0,i_1, i_2, ..., i_{j\!-\!1}}.
  \label{phi}
  \end{eqnarray}

%If $\{i_1, i_2, ..., i_{t}\}=\{\underbrace{0, 0, \cdots, 0}_{k-1},i_k,\underbrace{1, 1, \cdots, 1}_{t-k}\}$ $(i_k\neq 0)$, we find
% \begin{equation}
% N^{i_j}_{0,i_1, ..., i_{j\!-\!1}}\!=\! \!\left\{\!
%  \begin{array}{ll}
%\frac{(m\!+\!1)^t}{2}\!-\!(m\!+\!1)^{t\!-\!j} & 1\leq j<k\\%j\!=\!1,2,\cdots, k\!-\!1   \\
%   \frac{(m\!+\!1)^t\!-\!(m\!+\!1)^{t\!-\!k\!+\!1}}{m} & j=k  \\
%    \frac{(m\!-\!1)(m\!+\!1)^t+(m\!+\!1)^{t\!-\!k}}{m} & j>k
%  \end{array}
%  \right..
% \label{Nik}
% \end{equation}
 %where $\xi_k=\frac{(m\!-\!1)(m\!+\!1)^t+(m\!+\!1)^{t\!-\!k}}{m}$. %$(m\!+\!1)^t\!-\!(m\!+\!1)^{t\!-\!k}-(m\!+\!1)^t/m\!+\!(m\!+\!1)^{t\!-\!k \!+\!1}/m$.

Let $\{i_1, i_2, ..., i_{t}\}=\{\underbrace{0, 0, \cdots, 0}_{k-1},i_k,\underbrace{1, 1, \cdots, 1}_{t-k}\}$ $(i_k\neq 0)$ in Eq.~(\ref{phi}) and replace $N^{i_j}_{0,i_1, i_2, ..., i_{j\!-\!1}}$ from Eqs.~(\ref{N_00k})-(\ref{N_k11j}), we get
\begin{eqnarray}
&&\Phi({\underbrace{0, 0, \cdots, 0}_{k-1},i_k,\underbrace{1, 1, \cdots, 1}_{t-k} })\nonumber \\
%  &=&\sum_{j=k}^t3^{t\!-\!j}[2(m\!+\!1)^t\!-\!4(m\!+\!1)^{t\!-\!j}]\!-\!\frac{4}{m}3^{t\!-\!k}[(m\!+\!1)^t \nonumber \\
%   & &\!-\!(m\!+\!1)^{t\!-\!k+\!1}]-4\frac{(m\!-\!1)(m\!+\!1)^t+(m\!+\!1)^{t\!-\!k}}{m}\sum_{j=k\!+\!1}^t3^{t\!-\!j}\nonumber \\
  &=&(m\!+\!1)^t(3^{t\!-\!k\!+\!1}\!-\!1)-\frac{4}{3m\!+\!2}[(3m\!+\!3)^{t\!-\!k\!+\!1}\!-\!1] \nonumber \\
  &&-\frac{4}{m}3^{t\!-\!k}[(m\!+\!1)^t\!-\!(m\!+\!1)^{t\!-\!k+\!1}]\nonumber \\
   & &-2\frac{(m\!-\!1)(m\!+\!1)^t+(m\!+\!1)^{t\!-\!k}}{m}(3^{t\!-\!k}\!-\!1)\nonumber \\
  &=&(m\!+\!1)^t(3^{t\!-\!k}\!+\!1)(1-\frac{2}{m})+(3m\!+\!3)^{t\!-\!k}\nonumber \\
  && \times\frac{2m+4}{m(3m\!+\!2)} +\frac{2}{m}(m\!+\!1)^{t\!-\!k}+\frac{4}{3m\!+\!2}\nonumber \\
  &>&0.
\label{PhiS}
\end{eqnarray}
%Let $\{i_1, i_2, ..., i_{t}\}=\{\underbrace{0, 0, \cdots, 0}_{k-1},i_k,\underbrace{1, 1, \cdots, 1}_{t-k}\}$ $(i_k\neq 0)$ in Eq.~(\ref{Sub_TX}) and replace $\Phi({\underbrace{0, 0, \cdots, 0}_{k-1},i_k,\underbrace{1, 1, \cdots, 1}_{t-k} \}})$ from Eq.~(\ref{PhiS}), we obtain
Therefore, %for any $m\geq2$,
$T_{\{\underbrace{0, 0, \cdots, 0}_{k},i_k, \underbrace{1, 1, \cdots, 1}_{t-k}\}}-T_{\{0\}} > 0.$
\\

%\begin{references}

%\nocite{*}

\end{document}